\documentclass[twocolumn,showpacs,superscriptaddress,10pt,pra]{revtex4-1}
\usepackage{amsmath,amssymb,graphicx,float}

\begin{document}

\title{Position-controlled trapping of nanoparticles and quantum dots on a fiber taper}
\author{Ryusei Watanabe}
\author{Daiki Yamamoto}
\author{Mark Sadgrove$^*$}

\affiliation{Department of Physics, Faculty of Science, Tokyo University of Science, 1-3 Kagurazaka, Shinjuku-ku, Tokyo 162-8601, Japan}
\email{mark.sadgrove@rs.tus.ac.jp}

\begin{abstract}
We investigate numerically and experimentally the properties of a two color optical fiber taper trap, for which the evanescent field of the modes in the fiber taper give rise to a three-dimensional trapping potential. Experimentally, we use the technique to confine colloidal nanoparticles near the surface of an optical fiber taper, and show that the trapping position of the particles is adjustable by controlling the relative power of two modes in the fiber. We also demonstrate a proof of principle application by trapping quantum dots together with gold nanoparticles in a configuration where the trapping fields double as the excitation field for the quantum dots. This scheme will allow the positioning of quantum emitters in order to adjust coupling to resonators combined with the fiber taper.
\end{abstract}
\maketitle
\section{Introduction} 
The use of nanocrystal quantum emitters coupled to micro and nanophotonic resonators and waveguides to provide single photons for quantum technologies has become a standard part
of the quantum optics toolbox~\cite{aharonovich2016solid}. One of the challenges that remains in using such nano-emitters is positioning them relative to the dielectric structure of
nanophotonic devices. In particular, since the waveguiude or resonator mode intensity (and hence the coupling between an emitter and the mode) is typically defined by the dielectric structure itself,
the ability to adjust the position of an emitter relative to that structure is required to maximize or minimize interfacing with the nanophotonic device as necessary for the application at hand.
To this end, optical trapping of atoms near nano-optical devices has been achieved in several settings in recent years.
Trapping using both waveguide modes~\cite{Rausch2color,Kato} and side illumination techniques~\cite{doi:10.1126/science.1237125,PhysRevLett.123.213602,RauschWGM} can allow coupling between the atoms and cavity to be adjusted.

On the other hand, solid state quantum emitters in nanocrystals are often utilised in colloidal form. The manipulation of nano-sized particles in liquid environments using immsersed nanowaveguides is proceeding apace from its initial foundations~\cite{kawata,skelton}, with a number of advances in selectivity~\cite{SasakiTaper, SasakiCapillary, TaperTrap}, use of higher order modes and particle binding~\cite{SileHigherOrder,SileBinding}, angular momentum~\cite{GeorgiyAngular}, and manipulation of novel particles~\cite{SasakiCapillary,GeorgiyJanus} achieved in the past few years.
In particular, the potential for counterpropagating modes to halt particles moving along a waveguide has been noted before~\cite{lei2012bidirectional,SasakiTaper}, although only in the context of creating a
region where no axial force is present. In this case, diffusion is still possible over the waveguide length because no restoring force exists.
True trapping of particles at a specific position determined by the structure of a liquid immersed nano-optical device is still a challenge which requires a novel approach.

Here, we demonstrate the trapping of colloidal nano-particles at an adjustable position along a tapered fiber. Using counterpropagating modes in the fiber of different wavelengths, the axial force from the mode light pressure on particles near the fiber due to the evanescent field falls to zero at exactly one fiber diameter i.e. at one point along the taper, with a restoring force arising due to the difference in evanescent penetration depths of the mode. By changing the applied laser power, we show that the trapping position can be moved along the taper.

We have three specific aims in the present paper. i) Numerically investigate aspects of the two-colour taper trapping method. ii) Demonstrate and elucidate basic properties of the trap including the position-adjustible trapping mechanism which is unique to the taper trap setup. (Although the concept was introduced in Ref.~\cite{TaperTrap}, it was used there for a specific application and the unique nature of the trap was not probed). iii) Demonstrate an application where quantum dots are trapped together with metal nanoparticles near to the fiber taper.

In terms of significance, we note that although we do not resolve or infer the fiber diameter at which trapping occurs in the present work, due to our use of a tapered fiber, our results imply that we can trap emitters
at a \emph{specific fiber diameter} in which case, we can in principle trap them at a position which optimizes their coupling to the fiber.

\begin{figure}[ht]
	\centering
	\includegraphics[width=0.9\linewidth]{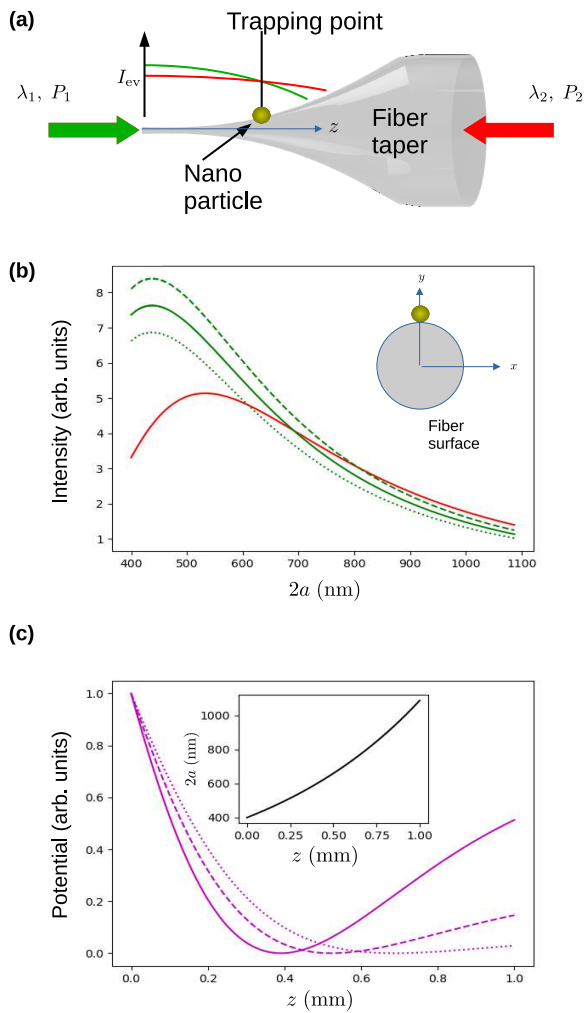}
		\caption{\label{fig:Concept}(a) Illustration of the experimental concept. Counterporpagating fiber modes with different wavelengths give rise to a trapping potential at a certain position along the fiber taper, due to the
		differing dependence of mode intensity on diameter. (b) Variation of the intensity crossover position for different relative mode powers.
		The results shown are for power ratios $R=P_{640}/P_{785}=0.9$ (green dotted line), $1.0$ (green solid line), and $1.1$ (green dashed line). The 785 nm mode intensity is shown by the red line. (c) Trapping potentials calculated for a 10 nm radius gold particle. The values of
		$R$ are 0.2 (solid line), 0.225 (dashed line), and 0.25 (dotted line). The potential minimum was set to zero in each case, after which the maximum values of each potential were scaled to unity to allow for easier comparison of the position of the trap minima.}
\end{figure}

\section{Physical principles and theory}
\subsection{Trapping potential}
The principle of the two-color taper trap technique is illustrated in Fig.~\ref{fig:Concept}(a). The evanescent field of the fundamental (HE$_{11}$) mode in the fiber taper region has a penetration depth into the surrounding medium (water in this research) which depends on its wavelength. Furthermore, the rate of attenuation of the evanescent field intensity $I_{\rm ev}$ as a function of the fiber diameter is faster for shorter wavelengths leading to a crossover point for the mode intensities which occurs at a certain fiber diameter. We note that this mechanism is related to the mechanism used in so-called two-colour traps for atoms using the evanescent portion of the modes of optical nanofibers~\cite{Fam2color, Rausch2color}.  However, in that case, it was necessary for the sign of the atomic polarizability to be different for each wavelength used in order to create a restoring force for the trap. In the present method, the restoring force arises due to the counter-propagating nature of the modes which leads to the dominant force changing direction about the point where the optical force vanishes.

The method used here is also reminiscient of the counter-propagating beam methods proposed and demonstrated by Ishihara and co-workers~\cite{SasakiTaper, Wada01}, although we note that mode behaviour in the taper region played no role in those studies, and thus trapping along the fiber-axis was not present.

In order to analyse this phenomenon quantitatively, it is useful to define the ratio $R=P_{\lambda_1} / P_{\lambda_2}$ which is the ratio of the power $P_{\lambda_1}$ in the shorter wavelength mode (wavelength $\lambda_1$) to that of the power $P_{\lambda_2}$ in the longer wavelength mode (wavelength $\lambda_2$).  Figure~\ref{fig:Concept}(b) shows how the crossover point changes for $R$ values of
$0.9$ (dotted green line), $1.0$ (solid green line) and $1.1$ (dashed green line). In all cases, the red line shows the intensity at the surface of the fiber for a mode with $\lambda_2=785$ nm and the green lines show the intensity at the fiber surface for a mode with $\lambda_1=640$ nm, as the fiber diameter is increased from 400 nm to 1000 nm. Here, we assume a quasi $y$-polarized fundamental mode (denoted HE$_{11}^y$) at a power of 1 mW for both wavelengths. The relevant axes are indicated in the inset of Fig.~\ref{fig:Concept}(b), and the position of the gold nanoparticle is assumed to be at the ``top" of the fiber as indicated in the inset, due to the fact that the radial potential from the gradient force is strongest there. (Due to the mode symmetry, there is also an identical trap position at the bottom of the fiber). For brevity, we do not reproduce the calculations of the mode fields here as they may be found in well-known textbooks~\cite{okamoto2006fundamentals}.

Although the basic trapping mechanism is well understood in terms of the crossover of mode intensities, in general, the power ratio required to achieve a taper trap at a given fiber diameter depends on the nature of the particle being trapped. In particular, for metal nanoparticles with a well defined plasmon resonance, the response to the shorter wavelength may be stronger leading to a zero-crossing of optical forces occurring at a larger diameter than the cross-over of optical intensities.

We explicitly demonstrate this in Fig.~\ref{fig:Concept}(c) for a spherical gold particle of radius 10 nm. This radius is sufficiently small compared to $\lambda_{1,2}$ that we may use the Rayleigh approximation.
Then, following Svoboda and Block, we write~\cite{Svoboda:94}
\begin{eqnarray}
F_{\rm scatt} &=& 8\pi^3\epsilon_mI_s\frac{|\alpha|^2}{3c\lambda^4},\nonumber\\
F_{\rm abs} & = & 2\pi\epsilon_mI_s \frac{{\rm Im}(\alpha)}{\lambda c}, \nonumber\\
F &=& F_{\rm scatt} + F_{\rm abs}\label{eq:theory}.
\end{eqnarray}
Here, $F_{\rm scatt}$ and $F_{\rm abs}$ are the $z-$directed scattering and absorption forces respectively, and $F$ is the total radiation pressure force at wavelength $\lambda$. $I_s$ is the mode intensity at the surface of the fiber taper, $\alpha$ is the nanoparticle polarizability, and $\epsilon_m$ is the relative electric permittivity of water.

Using the above expressions and the standard expression for the polarizability of a spherical particle (using wavelength dependent permittivities for gold from Johnson and Christy~\cite{JC}), we first calculate the total
force $F=F_{640} - F_{785}$, where the subscripts indicate the wavelength in nm, and then, using the relation\[U(z)=\int_{-\infty}^z F(z')dz',\] we numerically integrate the total force to give the potential shown in
Fig.~\ref{fig:Concept}(c) for three different $R$ values. Note that the minimum of the potential energy is adjusted to 0 and the maximum value is then scaled to 1 to allow easy discernment of the relative position of the trap minimum (i.e. the trapping position). The trapping position may be seen to increase as $R$ is increased.

Although the above analytical results are instructive regarding the basic behavior of the two-color taper trapping scheme, in our experiments the nanoparticle radius was 75 nm -  outside the regime where the Rayleigh approximation may be reasonably expected to hold. In this case, it is necessary to perform numerical simulations of Maxwell's equations in order to achieve results which may be compared to experimental measurements. We will discuss the relevant numerical simulations and their results in the following Section.

\subsection{Particle motion in the trap}
As we will show numerically in the following Section, the particle motion in the two-color taper trap turns out to be in the strongly over-damped regime, as with most optical traps in
fluids. If we approximate the potential as harmonic, then the motion of a particle in the trap is governed by the Langevin equation~\cite{bradac2018nanoscale, smart2018study}
\[
m\ddot{z} + \gamma\dot{z} + Sz = \sqrt{\Gamma}\zeta(t)
\]
where $m$ is the particle mass, $\gamma$ is the damping coefficient of the motion, $S$ is the trap stiffness, $\Gamma=2\gamma k_B T$ at room temperature $T$, and $\zeta(t)$ is a fluctuation term with units $\sqrt{\rm Hz}$ satisfying $\langle\zeta(t_1)\zeta(t_2)\rangle=\delta(t_1-t_2)$.
In the strongly overdamped regime, where $\gamma^2\gg 4Sm$, we can ignore the inertial term~\cite{smart2018study} and, assuming fluctuations can be neglected on the timescale of the particle entering the trap, we write
\[
\gamma\dot{z} + Sz = 0
\]
This equation has the solution
\begin{equation}
\label{eq:zeq}
z=A\exp\left(-\Lambda_+ t\right) + z_0,
\end{equation}
where $\Lambda_+ = S / \gamma$.
(Note that even with a non-negligible fluctuation term, the time averaged solution can be shown to be the same~\cite{smart2018study}).
This exponential relaxation to the trap center at $z_0$ is found to happen on the scale of seconds in the experiment.

\section{Numerical results}
\label{sec:Numerics}
\begin{figure*}[t]
	\centering
	\includegraphics[width=\linewidth]{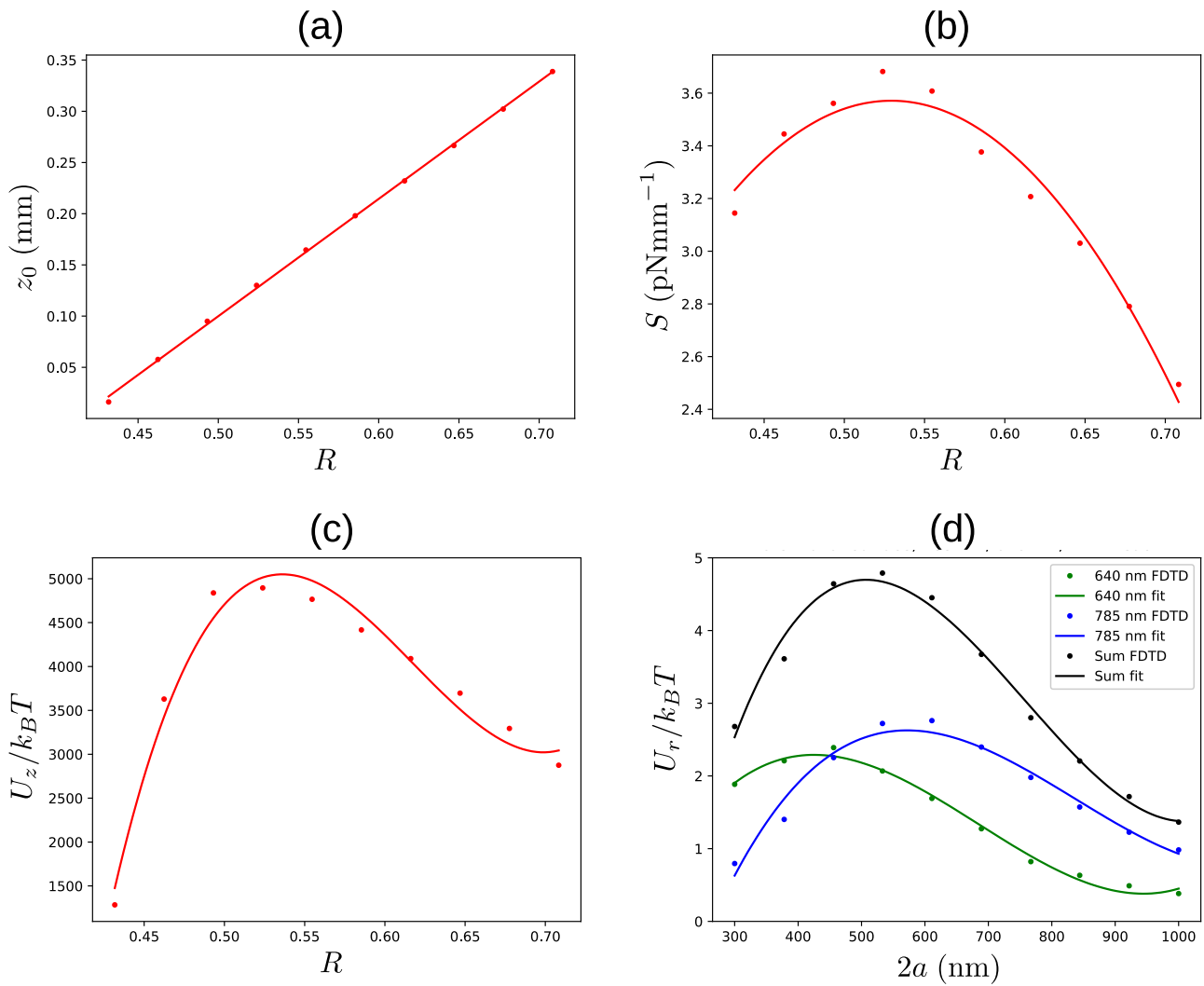}
		\caption{\label{fig:Numerics} FDTD simulation results for 1 mW in the longer wavelength mode. (a) Trap position as a function of mode power ratio $R$ for the 640 nm / 785 nm configuration (red points). The red line shows a linear fit to the data. (b) Trap stiffness as a function of $R$ (red points). The red line shows a second order polynominal fit to the data. (c) Trap depth $\pm 100\;\mu$m from the trap center as a function of $R$ (red points). The red curve shows a third order polynomial fit to the data.  (d) Radial trap depths as a function of fiber diameter $2a$ for 785 nm (blue points, blue curve) and 640 nm (green points, green curve) modes. In all cases, solid curves show third order polynomial fits to the data and are intended to guide the eye.}
\end{figure*}
For our numerical calculations, we used a commercial finite-difference time-domain (FDTD) Maxwell equation solver (Lumerical) and numerically evaluated the Maxwell stress tensor $\mathbf{T}$ using the fields calculated at the surface $\partial V$ of a cubical region $V$ encompassing the gold nanoparticle, but not the fiber. This allowed the optical force to be calculated by numerically evaluating the area integral
\[
\mathbf{F} = \oint_{\partial V} \mathbf{T}\cdot\mathbf{n}\;{\rm d}A,
\]
where $\mathbf{n}$ is the unit normal vector to the surface $\partial V$.
The gold nanoparticle was situated so that its surface was separated by 10 nm from the fiber surface. In our calculations, we assumed that the particle was trapped by the optical gradient force at the ``top" of the fiber (i.e. at a position $(x=0,\;y=a)$) (where the electric field of the quasi $y-$polarized fundamental mode is strongest). That is, the particle position was assumed to be as depicted in the inset of Fig.~\ref{fig:Concept}(b). We will discuss our assumption of quasi $y-$polarized fundamental modes later.

We performed simulations for a positive $z$ propagating mode with wavelength $\lambda_1=640$ nm and a negative $z$ propagating mode with wavelength $\lambda_2=785$ nm. In order to map simulations performed as a function of the taper diameter to results depending on position (which is what can be measured experimentally) it was necessary to assume a shape for the fiber given by $D_0\exp(z/L_0)$, where $D_0$ is the constant diameter of the waist region of the taper, and $L_0$ is the waist length. We used the nominal parameters $D_0=400$ nm, as measured
from scanning electron microscope observations of the fiber, and $L_0$=1 mm, which is the nominal value set during the fabrication process.

We first discuss our simulation results for the position $z_0$ of the trap minimum as a function of $R$, which is the result of principle interest in our present paper. The trap minimum was found directly from the force data by finding the power ratio which satisfied the condition $F_{\rm Tot} = F_{640}-F_{785}=0$ for a given diameter, and mapping that diameter to a distance $\Delta z$ from the start of the taper assuming an exponential taper shape~\cite{birks1992shape} as shown in the inset of  Fig.~\ref{fig:Concept}(c). The simulation results are shown in Fig.~\ref{fig:Numerics}(a) by red points. The solid line shows a linear fit to the data, where the gradient was found to be 1.1 mm per unit increase in $R$.
The linear variation of the trap position as $R$ changes is due to the approximately linear variation of the field intensity with diameter in the region considered. Roughly speaking, the trapping point is determined by the crossing point
of the intensities of each mode, and since these mode intensities show approximate linear dependence on $R$ so too does the trap position.

Note that the linearity of the positional change with $R$ is important for experimental checks of the taper trap. Due to the linearity, there is no need to know absolutely ``where" along the taper a particle is. Instead,
it is good enough merely to observe the change in position as $R$ is varied to check the model.

Next, we consider the trap stiffness $S$ as a function of $R$ which we also be calculated directly from numerical simulations of the force.
Specifically, we fitted a line to the points near the zero crossing of $F_{\rm Tot}$ whose gradient gives the trap stiffness at 1 mW of power in the longer wavelength beam. Results are shown in Fig.~\ref{fig:Numerics}(b).
Note that for the trap stiffness and trap potential depth results,
no analytical form exists for the results. We therefore chose to illustrate the trend in the data by fitting a polynomial to the data. We caution that these fitted curves are intended only to guide the eye,
and cannot be used to extrapolate the behavior of the simulated values outside the range of the data.
The trap stiffness is seen to peak at about 3.6 pN / mm before dropping off steadily as $R$ increases, and its average value is $\overline{S}\approx 3$ pN / mm. The qualitative behavior of $S$ is due to the fact that the surface intensity has a maximum at an optimum fiber diameter corresponding to $R\sim0.5$. For larger $R$, the trap position is pushed further up the taper where the
evanescent field of both modes becomes weaker, leading to lower trap depths and, accordingly, lower trap stiffness. Note that because the force is proportional to optical intensity and thus power, the trap stiffness $S'$ at a
different power $P'$ in mW can be found from the data shown in  Fig.~\ref{fig:Numerics}(b) at the same value of $R$ by calculating $S'=(P'/1$ mW$)S$.

The trap stiffness results also allow us to classify the trapping regime of the two-color taper trap. In particular, using Stokes' law, and ignoring the effect of the fiber, we can estimate the coefficient of damping to be $\gamma_0 = 6\pi\eta a_{\rm ns}$, where $\eta\approx 1\times10^{-3}$ Pa$\cdot$s is the viscocity of water at room temperature and $a_{\rm ns}=75$ nm is the radius of the nanoparticles, and calculating the particle mass as $m=\rho_{Au}\times 4/3\pi a_{\rm ns}^3$, where $\rho_{Au}=19300$ kg$\cdot$m$^{-3}$, we find $\gamma^2\approx1.6\times10^{-18}$kg$^2$s$^{-2}$$\gg 4Sm\approx5.2\times10^{-24}$ kg$^2$s$^{-2}$. Also, assuming the above value of $\gamma$, the diffusion term in the Langevin equation is found to be $\sqrt{\Gamma}=\sqrt{2\gamma k_B T}\approx 3.2\times10^{-15}$ kg$\cdot$m$\cdot$s$^{-3/2}$. Therefore, the trap is in the strongly overdamped regime, and we expect particles to move in an exponential trajectory towards the trap center without oscillating about it, as discussed in the preceding Section. It is important to note at this point that the true trapping potential is anharmonic because the evanescent mode intensity eventually falls to zero as the fiber diameter becomes large compared to the wavelength. This reduction in the trap potential relative to an ideal harmonic trap will lead to a reduction of the trap stiffness experienced by particles which enter the trap from outside the area approximately $\pm 100\;\mu$m of the trap center.

We also calculated the taper trap depth by numerically integrating the force over the $z-$axis to evaluate the potential energy. Because of the asymmetric and anharmonic shape of the trap far from its center, the trap depth is difficult to define. We approximated the trap depth as the average of the potential difference relative to the trap center at positions $\pm 100\;\mu$m about the trap center. We make the following observations about the  results which are shown in  Fig.~\ref{fig:Numerics}(c).
First, it is notable that the trap depth is large compared with $k_B T$ for room temperature $T$, rising to almost 5000 in $k_BT$ units. This is essentially because the trapping range is large - the restoring force works on the particle over hundreds of $\mu$m until it reaches the trap center. Second, we note that the trap depth falls when $R$ becomes large because as the power in the shorter wavelength mode increases, the particle is pushed to thicker and thicker parts of the taper, where the evanescent field intensity becomes less, leading to less overall force.

We used the FDTD simulations to calculate how the ``radial" trap depth changes along the $y$ axis as a function of fiber diameter as shown in Fig.~\ref{fig:Numerics}(d). To calculate the radial trap depth, we used separate FDTD simulations to those used to produce the results shown in Figs.~\ref{fig:Numerics}(a-c). In particular, the $y-$position of the particle was moved from a position distant from the fiber to a position 10 nm from the particle surface and the force calculated for each position.
The forces where then integrated over the $y$ axis to estimate the potential depth along this (radial) direction. We performed these simulations separately for 640 nm and 785 nm wavelength modes. The 640 nm mode trap depth peaks near a diameter of 420 nm while that for the 785 nm mode peaks near a fiber diameter of 570 nm. These diameters correspond to the effective diffraction limits at the two mode wavelengths for a fiber immersed in water. The total trap depth is given by the sum of the trap depths for the individual modes. It is notable that the trap depths for the radial trap and the taper trap along the $z-$axis differ by nearly three orders of magnitude. This is largely attributable to the very different range over which the forces act, with the radial gradient force being non-negligible only within one or two wavelengths of the fiber surface ($\lesssim 1\mu$m) while the
taper trap forces act over a range of hundreds of $\mu$m. For this reason, the trap lifetime is largely attributable to the radial trap depth. Above a fiber diameter of $1\;\mu$m, the radial trap depth may become too shallow to support long trapping times for reasonable input laser powers. Regarding the azimuthal trap depth, this requires a well defined polarization state, in order to be well defined itself, but experimentally we do not control the polarization. Therefore, consideration of the azimuthal trap potential is not particularly relevent to the results we consider in the present work, and we focus on the radial potential along the $y$ axis.

The above numerical exploration of the two-color taper trapping scheme characterizes its basic behavior. Note that in experiments, the relative trap position and trap stiffness can be measured in principle, while the trap depth cannot.

\section{Experiment}
\subsection{Experimental setup}
\label{sec:Expt}
\begin{figure*}[t]
	\centering
	\includegraphics[width=0.7\linewidth]{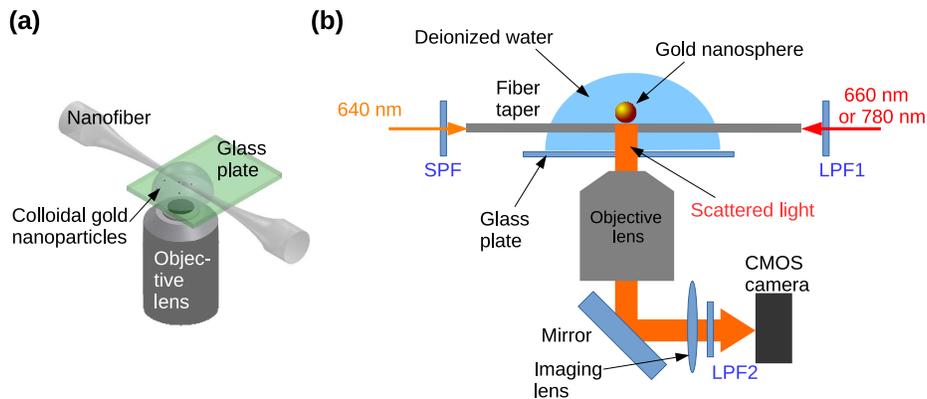}
		\caption{\label{fig:ExpSetup} Experimental setup.(a) Model of the principle elements of the setup showing the insertion of the waist region (nanofiber region) of a tapered optical fiber into a water droplet on a glass slide. Imaging takes place through the glass slide using an objective lens. (b) Detailed schematic diagram of the setup. Here, LPF - long pass filter, CMOS - complementary metal-oxide-semiconductor.}
\end{figure*}
We now move on to the experimental portion of this work, starting with the experimental setup. Figure~\ref{fig:ExpSetup}(a) shows a an overview of the main experimental elements. A tapered fiber (typical transmission $95\%$, typical waist diameter $\sim 400\pm$ 50 nm) is prepared from commercial optical fiber (Thorlabs 780HP) using a standard heat and pull method~\cite{OISTTaper} in a two stage pull.
The first stage produces a gradual taper which is adiabatic and leads to minimal loss through the core-mode cutoff regime. The second stage of the taper occurs in the few-mode regime where the adiabatic condition is easier to satisfy and  rapidly reduces the fiber diameter, producing a relatively steep taper gradient which is suitable for the taper trapping application considered here. Typical tapers produced by this method have a constant diameter waist region of only a few hundred micron in length. The taper satisfies the adiabatic condition for
all wavelengths used in the experiment, and since the input mode is the fundamental mode in all cases, we do not expect higher order modes to be present even in the thicker part of the taper where the single mode condition does not hold.

The fiber taper is introduced to a droplet of pure deionized water atop a glass slide. From below, the fiber can be imaged using a microscope objective (20x, Sigma Koki) lens followed by a 50 mm focal length plano convex imaging lens (``tube lens") in a homemade microscope setup. Insertion of the fiber into the droplet is achieved by moving the glass slide up towards the mounted fiber taper until the taper just touches the water surface. The taper is then sunk into the droplet by applying deionized water droplets from a micropipette directly above the fiber. Post-insertion, the transmission typically drops by between $0$ to $5\%$. After insertion into the droplet, colloidal gold nanospheres with no surface functionalization (Nanopartz A11-150-BARE-DIH) are introduced using a micropipette producing a concentration of approximately $10^5$ particles per $\mu$L.

In principle, the taper's constant diameter waist region  corresponds to a region of local maximal scattering intensity when the taper is viewed through the
objective lens. We note that while we are able to focus on this region of maximum intensity, our knowledge of the exact position along the taper, and therefore our ability to
reconstruct the fiber diameter at each point within our microscope observations, is not accurate,  due to uneven scattering. We would conservatively estimate at least $\pm 100\;\mu$m of uncertainty in the position of the fiber center. Due to the sharp gradient of our fiber near its center, this limits our ability to be sure about the absolute diameter of the taper at the position where trapping occurs. For this reason, we focus on the more easily observable parameter of \emph{relative} trap position in the current experiments. As discussed in the previous section, this is good enough to allow quantitative comparison with the numerical results.

Figure~\ref{fig:ExpSetup}(b) shows a more detailed schematic diagram of the experiment. Light of both wavelengths was generated by separate free-running diode lasers which were coupled to optical fibers.
The shorter wavelength mode ($\lambda_1=640$ nm) enters from the left end of the fiber through a 650 nm cutoff short pass filter SPF,
while the longer wavelength mode ($\lambda_2=785$ nm or $\lambda_2=660$ nm) enters from the right through a 650 nm long pass filter LPF1. The filters prevent light from the counterpropagating lasers from entering the
oppositely situated laser. Note that no optical isolators are used, and the lasers are unstabilized free-running diode lasers with bandwidths of order 1 nm.

Particles which are sufficiently close to the fiber taper surface can be trapped by the radial gradient force after which the are propelled along the fiber towards the center of the taper trap. Scattered light from such particles
is captured by the microscope objective and focused on the sensor of a CMOS camera (Thorlabs DC1545). Video recordings of this scattered light constitute the basic raw data of the experiment. However, only the light along the fiber axis is relevant to the experiment. We therefore extract the pixels along the axis of the fiber in each video frame and concatenate the results to form a reduced raw data set which displays the ``trajectories" of particles, being their $z$ position as a function of time.

\begin{figure*}[t]
	\centering
	\includegraphics[width=0.7\linewidth]{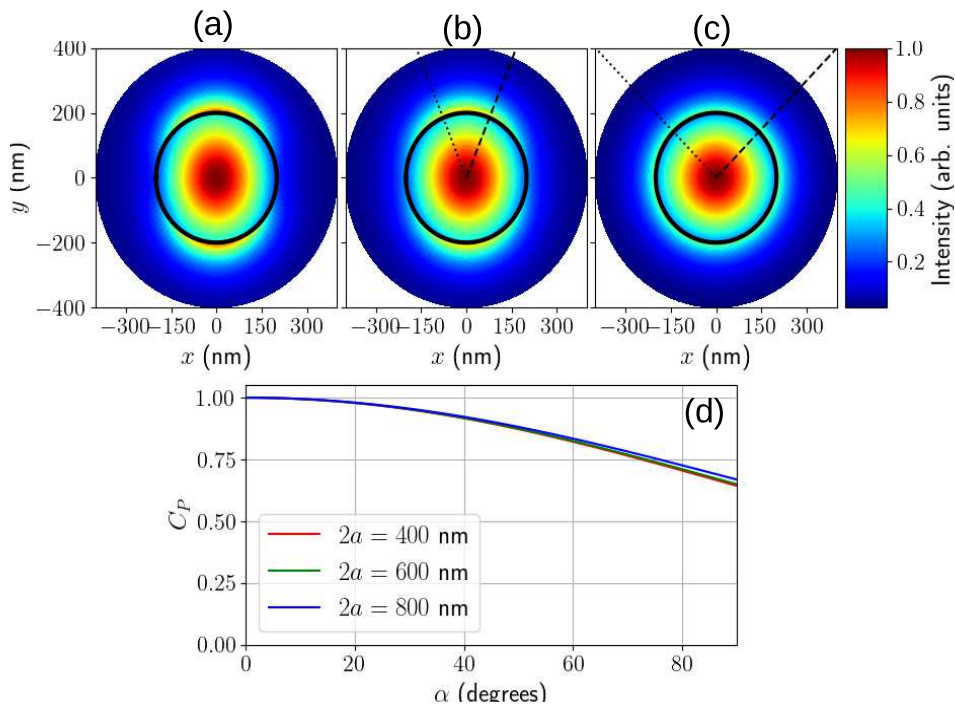}
	\caption{\label{fig:Pol} Calculations showing the effects of relative rotation of the fiber mode polarizations. In (a), the polarizations
	of the two modes are aligned and the power in each mode is the same. In (b) and (c), the relative rotations of the quasi-linear polarized fundamental modes are 45 and 90 degrees respectively. The polarization axes are shown by dotted and dashed black lines for the 785 nm and 640 nm wavelength mode respectively. The thick black circles indicate the fiber surface in each of (a)-(c).  In (d), the correction factor $C_P$ due to the  effect of relative rotation between the (linear) polarizations of the two modes is shown for several taper diameters $2a$ as indicated in the legend.}
\end{figure*}

\subsection{Non-ideal aspects of the experiment}
\label{sec:NonIdeal}
Let us now comment on some aspects of the experiment which are not ideal and the necessary corrections required to deal with them.

First, we chose not to use mode polarization control in this experiment. This means that the exact polarization state of the mode in the fiber taper region was unknown in our experiments. This is a deliberate choice; Controlling the mode polarization precisely requires significant extra equipment in excess of the simple setup illustrated in Fig.~\ref{fig:ExpSetup}(b). As we aim for this manipulation technique to be economical and simple, if polarization control were required it would significantly reduce its attractiveness. In a previous paper, little difference was found between experiments with a nominal level of polarization control and those with no control~\cite{TaperTrap}. We therefore expect that even without the polarization states matched to those used in simulations, a reasonable level of agreement between numerical predictions and experimental results should be found.

In Figs.~\ref{fig:Pol}(a), (b), (c), we show the effect of relative rotation between the (linear) polarization of the two fundamental modes by angles $\alpha=0$, 45 degrees and 90 degrees. It may be seen that the relative rotation
leads to an azimuthal broadening of the lobes at the top and bottom of the fiber, and indeed the two lobes merge when $\alpha=90$ degrees. Accompanying this change in the maximum intensity distribution is a drop in intensity of the
evanescent field at the fiber surface. This intensity $I_\alpha(x=0, y=a+a_{\rm ns})$ normalized to the $\alpha=0$ case gives a correction factor $C_P$ shown in Fig.~\ref{fig:Pol}(d), relative to the simulations for which the polarization was perfectly aligned. The correction factor is shown for several different taper diameters as indicated in the legend. The correction factor is seen to reach a minimum of $C_P\approx0.63$ when $\alpha=90$ degrees.
As may be seen, changing the fiber diameter does not produce a significant change in the behavior of $C_P$. The situation does not change significantly when elliptical polarizations are considered, although angular momentum can be transferred to the particle in this case. Because this is independent of the motion along the fiber axis, it is not expected to affect the results considered here.

Second, the velocity dependent damping force to the particles' motion is expected to be larger than that given by Stokes' law due to the presence of the fiber. Rigorous results do not exist for the effect of a cylindrical body
such as a fiber on the damping force. It is, therefore, typical to use corrections developed for a particle close to a plane~\cite{SileHigherOrder,goldman1967slow,krishnan1995inertial}. However, in this case, the distance
from the fiber to the particle surface must be known or introduced as a fitting parameter. Given that the particle diameter here is only a few times less than the fiber diameter, the use of these corrections may be dubious
in our case. We have therefore chosen to calculate the coefficient of damping $\gamma$ by measuring the terminal velocity $v_t$ of particles transported in the presence of only one mode,
and using the relation $\gamma = F_z/v_t$, where $F_z$ is the optical force on the particle along the $z$ axis due to the mode. Since the optical force cannot be directly measured experimentally,
we use the FDTD simulation value of $F_z$ in our calculation.

Finally, we mention non-ideal aspects of the experiment caused by fiber-particle interactions. Because our experiment uses a simple method for immersing the fiber taper in a droplet, standard water surface tension is required, and surfactants cannot easily be used. For this reason, some particle sticking to the fiber is inevitable. We assumed that our experimental results were not affected by sticking so long as the following conditions were met: i) Particle adhesion to the fiber did not significantly decrease the transmitted power and ii) the position at which particle sticking occurred was outside the region where particle trapping was observed.
The data used for this paper satisfied these conditions. Nonetheless, it was necessary to perform experiments within a brief time range in order to reduce the chance of a particle sticking in the region to be measured.
We consider trajectories measured over 10 s intervals for various values of $R$. Because of the non-deterministic nature of particles entering the trapping region and the limited time frame for each measurement, the number of trajectories observed
unavoidably varied for each value of $R$.

\subsection{Data analysis}
\label{sec:DataAnalysis}
In this subsection we briefly describe the analysis methods used on the data.

Regarding the detection of transporting trajectories in the data, we first use a peak detection method on each frame of the data, and then correlate nearby peaks to establish trajectories~\cite{TaperTrap}.
This method works reasonably well when only transporting trajectories are present, and the density of trajectories is not too high. It works less reliably in the case where persistent trapping trajectories are present,
due to ambiguity over which of multiple peaks in a frame belong to which unique trajectory.
Once trajectories have been detected, they can be fitted by Eq.~\ref{eq:zeq} allowing the extraction of the relaxation rate parameter $\Lambda_+$. From here, we can calculate the experimental trap stiffness $S$ if the
damping coefficient $\gamma$ is known.

In the case of trapped particle trajectories, a more robust method of detection is necessary. We used a modified Hough-like method~\cite{hough} to detect trapped particle trajectories as follows.
Algorithmically, we scan a horizontal line over the data in the direction where trajectories are found. Points within $\pm 2\Delta$ of the line's vertical position are considered to be part of a trapped trajectory, where $\Delta$
is the observed trajectory width, which depends on the imaging optics, and which is determined in the following Section.
The vertical position of the line which leads to the most points in the trajectory is considered to be the center of the trapping trajectory and thus the trapping position for particles. Although not completely independent
of the parameters used to find the peaks in the data, this method gives a way to objectively assign a trapping position to data for which trapping trajectories are observed.

\section{Experimental results}
\subsection{Evaluation of $\gamma$}
\label{sec:gamma}
\begin{figure}[t]
	\centering
	\includegraphics[width=1.0\linewidth]{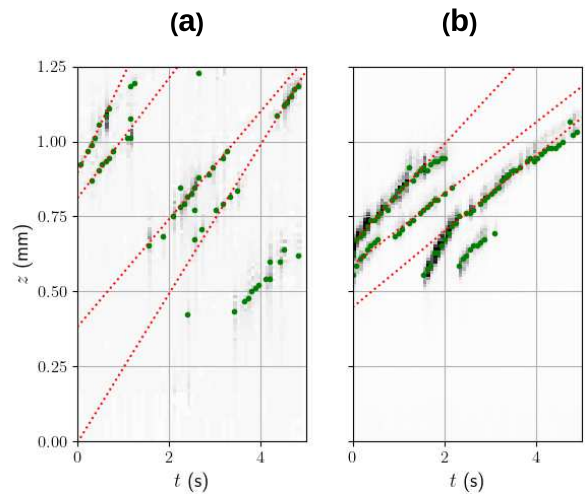}
		\caption{\label{fig:Expt} (a) shows raw data exhibiting transporting trajectories for an experiment where 9 mW of 660 nm wavelength light was present in the fiber taper.
		(b) shows raw data for the case where 8 mW of the 785 nm mode was present in the fiber taper. In both cases, the detected peaks in the data are shown by green circles to
		enhance the visibility of the trajectories. Red dotted lines show fits to trajectories which started at $z>0.75$ mm.}
\end{figure}
Examples of raw data from the experiment when only one mode is present in the fiber taper are shown in  Fig.~\ref{fig:Expt}(a) and (b). In the figures, we overlaid detected peaks in the data as green circles to make the trajectories clearer. In (a), only the positive $z$ propagating 660 nm mode (power 9 mW) is present and in (b), only the positive $z$ propagating 785 nm mode (power 8 mW) is present and in both cases only transporting trajectories are seen. This allows
us to calculate the velocity of the particle and estimate the damping coefficient $\gamma$ by comparing the results to simulations. Note that by performing the same calculation for two separate wavelengths 660 nm and 785 nm (the difference in simulated optical force for 660 nm and 640 nm is neglible, so we only consider the 660 nm mode here), we can ascertain whether other effects are at play, as ideally $\gamma$ should be independent of wavelength.
Here, the choice of where to fit a straight line has an impact on the results.
Tapered fibers have a constant diameter region at their very center (which, when of sub-micron dimension is referred to as a ``nanofiber") where the evanescent field intensity and thus optical force are constant. In this region, the particle velocity rapidly reaches a constant value due to the velocity dependent dampling force. After the taper begins, the evanescent field reduces, and the force reduces, leading to a change in velocity. The exact nature of the velocity change is somewhat complicated by the fact that the decreasing radial gradient may allow the particle to move away from the fiber surface and thus encounter less resistance. We do not account for this effect here.

In Fig.~\ref{fig:Expt}(b), which is from the same data set as Fig.~\ref{fig:Expt0}, straight line trajectories were seen for $z$ above 0.75 mm. Below 0.75 mm the trajectories have clearly started to curve, suggesting this region of the fiber is tapered. However, the exact point where the taper begins between 0.75 mm and 1 mm  is
not possible to ascertain. We conservatively chose to fit lines to detected trajectories with starting points above $z=0.75$ mm. For the 660 nm wavelength mode, we found an average velocity of $v=237\pm54\;\mu$ms$^{-1}$, and for the 785 nm
wavelength mode the average velocity was $v=137\pm20\;\mu$ms$^{-1}$. The FDTD calculated optical forces for a fiber diameter of 400 nm were 3.89 pN and 1.45 pN for 660 nm and 785 nm wavelength modes respectively.
We calculated the values of $\gamma$ at each wavelength to be $\gamma_{660} = 16\pm 3.7$ pN$\cdot$m$^{-1}$s, and $\gamma_{785} = 11\pm 1.6$ pN$\cdot$m$^{-1}$s. Although it is not possible to rule out effects other than
damping contributing to these results, we note that these values overlap within their respective one standard deviation errors as would be expected if damping alone gave rise to the terminal velocities of the particles.
We take the mean of these two values as our value of the damping constant for use in the calculation of trap stiffness later, i.e. $\gamma=13\pm 2$ pN$\cdot$m$^{-1}$s.

We emphasize again that for these calculations, we assumed that effects due to particle-particle interactions, and thermal effects (convection, viscocity change, etc) are negligible. All differences were attributed to differing values of $\gamma$.

\subsection{Qualitative properties of taper trapping}
\begin{figure}[t]
	\centering
	\includegraphics[width=1.0\linewidth]{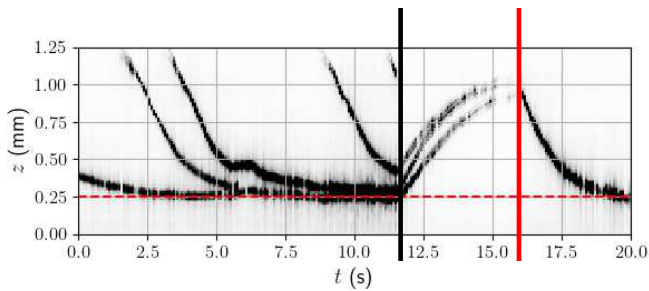}
		\caption{\label{fig:Expt0} Sample raw data from an  experiment where 8 mW of the 785 nm mode and 4 mW of the 640 nm mode are counter-propagating within the fiber.
		The horizontal red dashed line shows the estimated position of the trap. The vertical black line shows the time at which the 640 nm laser was blocked.
		The vertical red line shows the time at which the 640 nm laser was restored.}
\end{figure}

In Fig.~\ref{fig:Expt0}, we show raw data from an experiment where counterpropagating 785 nm wavelength and 640 nm wavelength modes were present in the fiber taper at powers of 8 mW and 4 mW respectively.
These data reveal a number of qualitative characteristics of the experiment which we now comment on. First, we describe the overall nature of the trajectories seen. For this ratio, particles are loaded into the trap by travelling in the negative $z-$ direction to the trap position at $\sim0.25$ mm. The 640 nm light propagates in the negative $z$ direction while the 785 nm light propagates in the positive $z$ direction. In the experiment for the data shown, the 640 nm wavelength mode was blocked at about 11 s and re-introduced at about 16 s, as shown by vertical black and red lines  respectively. Our first observation regarding the data is that multiple particles may enter the trap. For nanoparticles of the size used, the nanoparticles remain in suspension over the duration of the experiment, and there is no way to deterministically control the number of particles which enter the trap. The average particle number can be non-deterministically adjusted by diluting the nanoparticle solution, with the caveat that this can make trapping events rare and data hard to procure. Our observations suggest that multiple particles being in the trap do not \emph{significantly} alter the trap position or other discernible trap properties at least for the a particle number $n\lesssim 5$ which was typically seen in our experiments. Nonetheless, small changes are possible, as suggested in the data at time $\sim 6$ s for example. The presence of multiple particles also makes estimates of the trap lifetime difficult in the current experiments. We can only say that qualitatively, the traps are long lived with lifetimes on the scale of tens of seconds, which is not in conflict with the deep potential wells found by FDTD simulation of the taper-trap. Second, we note that the observed trajectories are in agreement with the prediction of an over-damped trap, with particles relaxing in an approximately exponential trajectory to the trapping point without oscillating about it. Third, we note that our resolution of the position of the trapped particles, as determined from the standard deviation from a Gaussian fit to the trajectory cross-section, is $\Delta = 23 \mu$m.
This is considered to be the minimum uncertainty in our measurements of the trap position.

\subsection{Trapping of gold nanoparticles as a function of $R$}
\begin{figure}[t]
	\centering
	\includegraphics[width=1.0\linewidth]{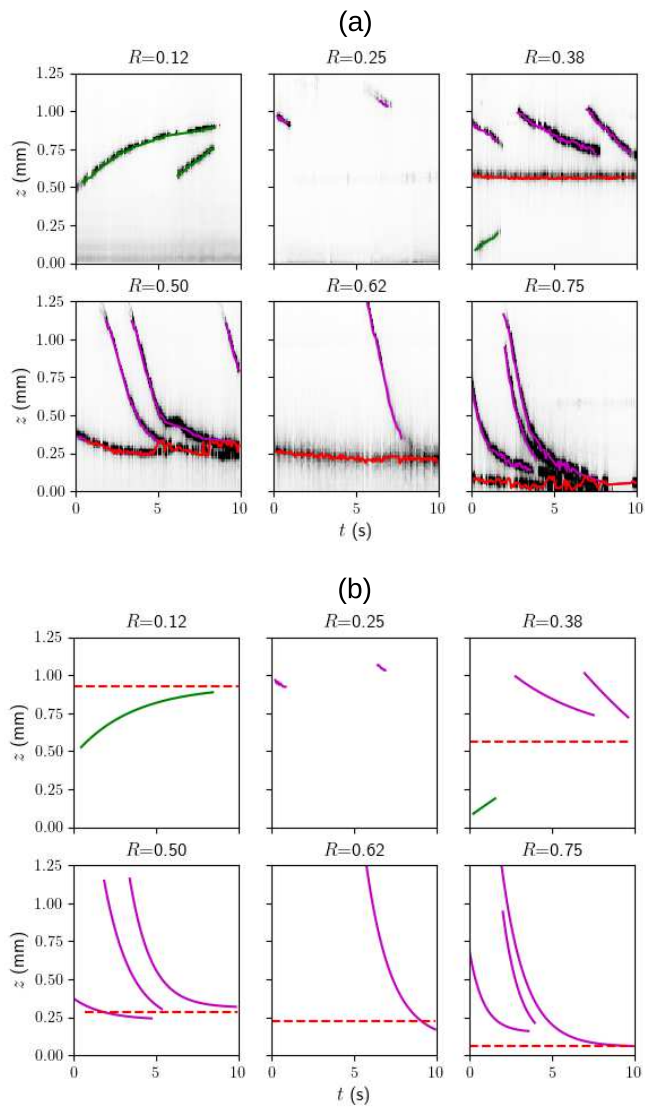}

		\caption{\label{fig:Expt1} (a) Sample trajectories for the 640 nm / 785 nm configuration as a function of $R$ (value shown above each panel). Black tracks show raw data, with detected trajectories overlayed as purple curves (negative direction transporting trajectories), green curves (positive direction transporting trajectories) and red curves (trapped trajectories). The position of the trapping trajectories may be seen to move as the power ratio changes. (b) Exponential fits to detected trajectories in (a) (magenta and green curves) are shown along with the mean position of detected trapping trajectories (red dashed lines). }
\end{figure}

We now move on to our principle experimental results. Our main aim in the experimental investigations was to confirm the unique characteristic of the taper trap scheme - the ability to control the trap position by varying $R$.
Experimentally, this was achieved by fixing the power of the longer wavelength and adjusting the power of the 640 nm mode. At least 10 s of video data was taken at each different 640 nm mode power. The trapping position
was identified using the Hough-like method described in the previous Section. Data sets are shown for each $R$ value in Fig.~\ref{fig:Expt1}(a), with the values of $R$ shown above each panel. Clear trapping was seen for the cases
$R=0.38$ through $R=0.75$, with a clear variation in the position of the horizontal trapping trajectories seen. In addition, although a horizontal trajectory was not seen in the case of $R=0.12$, the longer of the two trajectories observed clearly becomes nearly horizontal near its end-point, signifying the onset of trapping. Although such nonlinear behavior can also arise due to the taper itself (i.e. the field weakens as the fiber thickens leading to the particle slowing down), this is not the reason for the flattening of the trajectory, as may be seen by the fact that for $R=0.25$ and higher, straight-line trajectories are seen in this region. We therefore used a fit of Eq.~\ref{eq:zeq} to the $R=0.12$ data
to extract the trap position $z_0$ and used this value in our analysis of the movement of the trapping position. On the other hand, we note that for $R=0.25$, no significant trajectories were seen during the allotted detection time.
This sometimes occurs due to the non-deterministic nature of the experiment, i.e., sometimes there are not sufficient numbers of particles near to the fiber for trajectories to be observed.

Qualitatively, it is notable that at smaller $R$, particles can approach the trap from both sides, while as $R$ increases, particles only enter with negative velocities. This is due to the fact that as the trapping position moves up the taper
due to stronger 640 nm light, it is more likely for a particle to be radially trapped in the thinner region of the fiber, where the evaneseent field intensity is higher, and then transported to the trap position.

In Fig.~\ref{fig:Expt1}(b), we show exponential fits to transporting trajectories for each value of $R$ (magenta and green lines for negative and positive transporting trajectories respectively) along with detected
trapping positions (dashed red lines). These exponential fits allow us to extract $\Lambda_+$ from which the trap stiffness may also be calculated.

\begin{figure*}[t]
	\centering
	\includegraphics[width=1.0\linewidth]{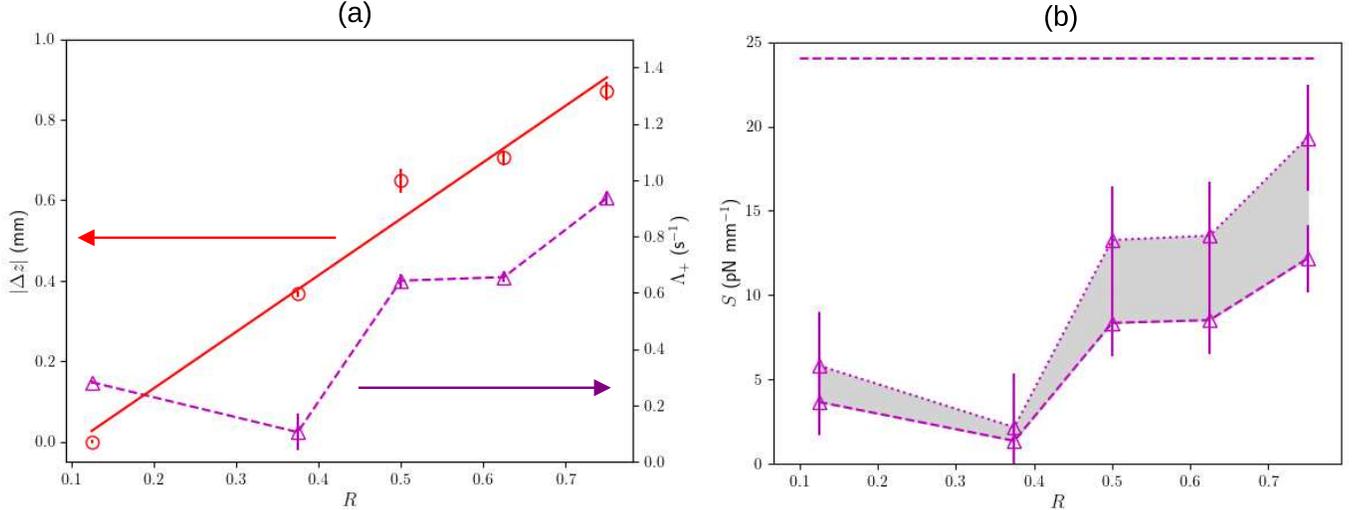}
	\caption{\label{fig:Expt1b} (a) Red points show detected trapping positions at each value of $R$. In the $R=0.12$ case only the trap position was estimated from the fitted value of $z_0$ of the positive transporting trajectory seen for that $R$ value in Fig.~\ref{fig:Expt1}(b). In all other cases, the values come from the mean of the detected trapping trajectories shown by red dashed lines in Fig.~\ref{fig:Expt1}(b). The red line shows a linear fit to the data.
	The gradient of the fitted line is $1.4\pm0.1$ mm per unit change in $R$. Magenta points connected by a broken line show the value of $\Lambda_+$ each $R$ value.
	(b) Magenta points connected by a broken line show the calculated value of $S$ at each $R$ value, whereas the points connected by a dotted line show the same data multiplied by $C_P$ to show the maximum correction due to polarization misalignment. The grey region between the two curves shows where the true value is expected to lie. In all cases, error bars show $\pm$ one standard deviation. The dashed horizontal line shows the approximate mean value $S'$ of the trap stiffness predicted by simulations. }
\end{figure*}
The experimentally measured values of the difference in trap position $|\Delta z|$ as $R$ was varied are shown in Fig.~\ref{fig:Expt1b}(a) as red points. Here, the difference is taken with respect to the value found in the $R=0.12$ case.
The error bars show $\pm 1$ standard deviation of the $z$ positions in the trajectory. A fit to the position data is shown by the solid red line.
The gradient of the fit is $1.4\pm0.1$ mm per unit change in $R$. We note that this value is close to the prediction of simulations of 1.1 mm per unit change in $R$.
Purple triangles in Fig.~\ref{fig:Expt1b}(a) show values of $\Lambda_+$ extracted from fits to the trajectories at each value of $R$.

In Fig.~\ref{fig:Expt1b}(b), we plot calculated values of the trap stiffness $S$  as purple triangles with a dashed magenta line connecting the points to guide the eye. We also include points scaled by the correction factor $C_P$ and join them by a dotted line. The shaded region in between is where the true value of $S$ is expected to lie. The dashed magenta line shows the expected mean value of the trap stiffness $S' = 8 \overline{S}=24$ pN / mm. Error bars are calculated from the error in $\gamma$ (see Section~\ref{sec:gamma}) and show one standard deviation.

The maximum value of $S$ is found to be $\approx 20$ pN / mm. This is still less than the average value expected for experiments where the 785 nm mode power is 8 mW (horizontal dashed line in Fig.~\ref{fig:Expt1b}(b)), suggesting that other effects, including the anharmonicity of the trap, may be non-negligible. We note that the qualitative form of the variation in $S$ with $R$ also does not match the numerical predictions, although this is not surprising, since, unlike the case for $\Delta z$, matching would require exact knowledge of the local taper diameter in the experiments.

\subsection{Transport and trapping of quantum dots adhered to gold nanoparticles}
Our second experimental results used a variation of the two-color taper trap considered so far. Our goal was to trap quantum dots adhered to gold nanoparticles close to the fiber taper using a taper trap, with the trapping modes also serving as excitation light for the quantum dots (Invitrogen Qdot 800 ITK). Because the quantum dots we used have a broad emission spectrum between 750 nm and 800 nm, we chose the long wavelength mode to have a wavelength of $\lambda_2=660$ nm for this experiment. This allowed the filtering of light scattered from the modes by the gold nanoparticles so that only fluorescence from the quantum dots was measured by the CMOS camera.

We note that the use of relatively close wavelengths for the two modes makes the trap position more sensitive to power fluctuations because both wavelengths are close to the plasmon resonance of the gold particles. We also note that for this reason, modelling the system with FDTD is more difficult as it requires a more exact match between the
model of the gold nanoparticle polarizability and the true experimental value. Nonetheless, the qualitative behaviour of the trap is similar to that seen for the 785 nm - 640 nm configuration, with the main
difference being that the ratio where trapping occurs is close to $R=1$.

\begin{figure}[htb]
	\centering
	\includegraphics[width=1.0\linewidth]{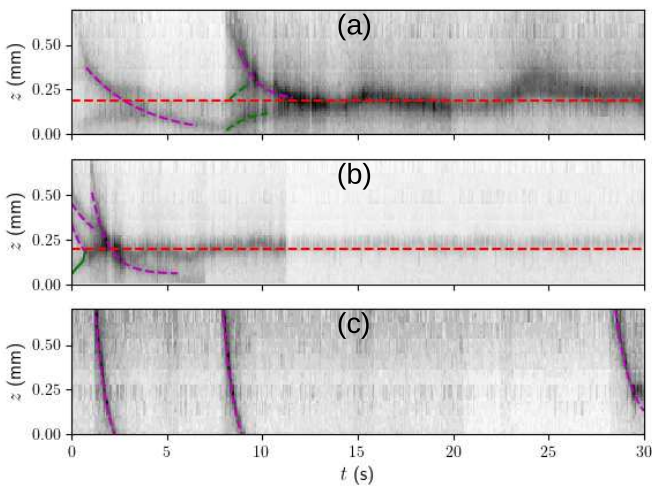}

		\caption{\label{fig:Expt2} Experimental results for the 640 nm / 660 nm configuration. Here, both gold nanoparticles and quantum dots are present in the solution. The values of $R$ are 0.9, 1.1 and 1.2 for panels (a), (b) and (c)  respectively. Exponential fits to detected negative transporting trajectories and positive transporting trajectories are shown by magenta and green curves respectively. The mean position of detected
		trapping tajectories is shown by a red dashed line in the $R=0.9$ and $1.1$ cases. Note that for $R=0.9$ and $R=1.1$, trapped trajectories were observed while at $R=1.2$, only tansporting trajectories were seen.}
\end{figure}

In Fig.~\ref{fig:Expt2} we show the results for the case when both gold nanoparticles and quantum dots are present in the solution. Trajectories were detected with a 750 nm long pass
filter installed before the CMOS camera so that only the quantum dot fluorescence was recorded. The three values of $R$ used were 0.9, 1.1, and 1.2 for panels (a), (b) and (c) respectively.
The detected fluorescence signal was much weaker than the scattered light signal, leading to a much reduced signal-to-noise ratio for this data
even at the most sensitive settings for our camera. Raw data with overlaid trajectory fits is shown in Fig.~\ref{fig:Expt2}.
As before, overlaid purple curves show detected negative directed trajectories, green curves show positive directed trajectories and red horizontal lines show trapping trajectories.

Trapping trajectories were detected for both $R=0.9$, with a mean position $z_{\rm min}=190\pm30\;\mu$m and $R=1.1$ with a mean position $z_{\rm min}=200\pm20\;\mu$m. The variance is clearly larger in the $R=0.9$ case, where the trapped trajectory drifted widely about the mean value. For $R=1.1$ a relatively stable trapping position is observed. This likely occurs due to the fact that for $R=0.9$ a zero force region can exist over the entire $\sim 0.5$ mm long constant diameter waist region of the taper. Here there is no restoring force, so the particles may drift due to Browninan motion.
For $R=1.1$, a true taper trap is expected to exist leading to a more stable trapping position.

We also extracted the values of $S$ in each case from fits to the observed trajectories. The calculated values were $S=8\pm2$ pN mm$^{-1}$ for $R=0.9$, $S=17\pm3$ pN mm$^{-1}$ for $R=1.1$.
It may be seen that for $R=1.2$ no trapping was observed in the observation window, and it is possible that the taper trap ceased to exist because the trap position was pushed out of the region where
a sufficiently strong radial evanescent field gradient force exists.

\section{Discussion and conclusion}
In this work, we first numerically demonstrated the ability to vary the trap position of a particle along a fiber taper, and systematically evaluated the characteristics of the two-color taper trap, including trap stiffness and trap depth. The strongly over-damped nature of the trap was also confirmed numerically. Experimentally, we confirmed the ability to move the trap position by adjusting the power ratio, and observed a dependence of the relative trap position on $R$ that was in good agreement with numerics. We also measured the trap stiffness, $S$ and found it to be smaller than
the expected value even after allowing for uncertainties in polarization. The most likely explanation is trap anharmonicity - the trap shape is only approximately parabolic over a range of about $\pm 200\;\mu$m about the trap center, but trajectories were observed over scales of $\sim 1$ mm. However, without the development of sophisticated dynamical simulations of the particle motion,
it is difficult to quantitatively evaluate the effect of the trap shape.

Next, by adjusting the longer wavelength mode's wavelength to a value outside the emission spectral range of a quantum dot, we were able to use the method to trap quantum dots adhered to gold nanoparticles near to the fiber taper.
Regarding this last experiment, it is important to consider another possible interpretation of the data. Because we mixed quantum dots and gold nanoparticles in this experiment, quantum dots were able to adhere to the fiber itself. This raises the possibility that rather than movement of quantum dots adhered to gold nanoparticles, the trajectories we observed were just those of gold nanoparticles which enhanced the fluorescence from quantum dots adhered to the fiber~\cite{sugawara2022plasmon,shafi2022bright} as they passed over the position of the quantum dots. We were able to rule out this interpretation by performing the experiments with gold nanoparticles with a different surface functionalization (Nanopartz A11-150-CIT) which did not lead to quantum dot - gold nanoparticle conglomerates. In this case, although gold nanoparticles were observed to be transported and trapped when scattered light was observed, no trajectories were observed when the long pass filter was inserted and only fluorescence from the QDs was detected.

Additionally, note that at present, we have no estimate for the number of quantum dots adhered to the gold nanoparticles. In terms of surface area, the gold nanoparticles are hundreds of times larger than quantum dots. Assuming that quantum dots adhere to the gold nanoparticles, but not to each other, it is possible that many tens of quantum dots are attached to each particle in the current experiments. At present, we believe it is more pertinent to focus on the issue of coupling fluorescence from trapped emitters to the fiber. Once this is achieved, efforts to reduce the number of emitters present may lead to novel single photon sources using this technique.

Another point to consider is whether the taper trap effect can really be useful for creating interfaces between quantum emitters and nano-photonic devices. The positioning demonstrated here is only accurate to the level of tens of $\mu$m, whereas state-of-the art fabrication techniques can achieve nano-positioning, and optical trapping of atoms near nano-devices achieves trap sizes below 1 $\mu$m. The reason we believe our technique is of interest despite these ostensibly unfavourable comparisons is that it is already useful to be able to trap a quantum emitter at a \emph{diameter} dependent position on a waveguide. The reason for this is that emitter-waveguide coupling depends sensitively on the waveguide diameter (as demonstrated for nanofibers in Refs.~\cite{Chandra, TUSSem}). Therefore, having a trapping technique which allows trapping at a given fiber taper diameter allows maximization of emitter-fiber coupling. Another situation where the resolution of the current technique might be sufficient is where a resonator couples to a fiber taper at a given point, such as in the case of whispering gallery mode resonators. In this case, our method could be used to trap an emitter at the region where the resonator couples to the fiber taper. In the situations given above, positioning at the level of hundreds of nm, as necessary for maximizing coupling to a photonic crystal cavity is simply not necessary and sufficiently good positioning can be achieved with a taper trap.

In conclusion, we have established numerically and experimentally the basic and unique properties of two-color optical fiber taper traps and used the technique to trap quantum dot - gold nanoparticle conglomerate particles near to a fiber taper. We anticipate that the taper trapping technique can provide a simple and flexible technique for coupling emitters and nanowaveguides optimally in a reconfigurable and reversible way.

\acknowledgements
This work was supported by the Nano-Quantum Information Research Division of Tokyo University of Science and by a JSPS KAKENHI Grant-in-Aid for Transformative Research Areas (grant number JP22H051351).

Data sets and computer programs used to create the figures for this paper are available at the following doi: \begin{verbatim} 10.6084/m9.figshare.22283731 \end{verbatim}.

\bibliography{QDGNS}

\end{document}